\documentclass[journal=jpccck,manuscript=article]{achemso}

\usepackage{amsmath,amssymb}
\usepackage{graphicx}
\usepackage{bm}
\usepackage{dsfont}
\usepackage{achemso}

\usepackage{float}
\usepackage{color,soul}
\usepackage{amsmath}
\definecolor{rojo}{rgb}{1,0,0}
\definecolor{verde}{rgb}{0,0.8,0.5}
\definecolor{azul}{rgb}{0,0,1}
\definecolor{rosa}{cmyk}{0,1,0,0}
\usepackage[toc,page]{appendix}
\usepackage{latexsym}
\usepackage{amsmath}
\usepackage{mathrsfs}
\usepackage{bm}
\usepackage{soul}
\usepackage{amsthm}
\usepackage{physics}
\usepackage{adjustbox}
\usepackage{bbm}
\usepackage{amsfonts}
\usepackage{amssymb}
\usepackage{color}
\usepackage{epsfig}
\usepackage{hyperref}
\hypersetup{colorlinks=true, citecolor=blue}
\usepackage{soul}
\usepackage[normalem]{ulem}
\usepackage[cmtip,all]{xy}
\usepackage{comment}
\usepackage{multirow}
\usepackage{mathtools}

\newcommand{\longsquiggly}{\xymatrix{{}\ar@{~>}[r]&{}}}

\makeatletter
\newcommand*\AuthorLabel[1]{%
  \expandafter\gdef\csname @authorextra@\@roman\acs@author@cnt\endcsname{#1}%
}

\newcommand{\corr}[1]{\textcolor{black}{#1}}

\author{Massimiliano Di Ventra}
\AuthorLabel{a} 
\email{diventra@physics.ucsd.edu}
\affiliation[UCS] {Department of Physics, University of California San Diego, La Jolla, CA, 92093, USA}

\author{Rafael Gutierrez}
\AuthorLabel{a}
\email{rafael.gutierrez@tu-dresden.de}
\affiliation[TUD]{Institute for Materials Science and Max Bergmann Center of Biomaterials, TU Dresden, 01062 Dresden, Germany}

\author{Gianaurelio Cuniberti}
\email{gianaurelio.cuniberti@tu-dresden.de}
\affiliation[TUD]{Institute for Materials Science and Max Bergmann Center of Biomaterials, TU Dresden, 01062 Dresden, Germany}
\alsoaffiliation[TUD1]{Dresden Center for Computational Materials Science (DCMS), TU Dresden, 01062 Dresden, Germany}

\title{Chirality-induced Spin-Orbit Coupling and Spin Selectivity}

\begin{document}


\begin{abstract}
We show that a spinor traveling along a one-dimensional helical path develops a spin-orbit coupling as a result of the curvature of the path. 
We  estimate the magnitude of the associated spin polarization and obtain values typical of many helical molecular structures that showcase the Chirality-induced Spin Selectivity (CISS) effect.
We find that this chirality-induced spin-orbit coupling ($\chi$-SOC), in conjunction with broken time-reversal symmetry, may be an important ingredient for the microscopic underpinning of the CISS phenomenon.
\end{abstract}

\maketitle

\noindent
\section{INTRODUCTION}
\label{intro}
Spin-orbit coupling (SOC) is a fundamental relativistic phenomenon arising from the coupling between the spin and orbital degrees of freedom of a spinful particle~\cite{Messiah}. In atomic systems, SOC scales as $Z^4$ with $Z$ being the atomic number; thus, SOC is typically smaller the lighter the atoms. 
 The spin itself can also be manipulated by coupling it to a magnetic field. 
 Therefore, it was a complete surprise that chiral molecules primarily 
made of organic elements, hence with weak SOC, display a spin response (selectivity) in the absence of any external magnetic field \cite{Ray1999,MD2011,Gohler2011,doi:10.1021/jacs.8b08421,doi:10.1021/jp509974z,doi:10.1021/nl2021637,Nogues2011,Kiran2016,Malajovich2000,Min2003,Wei2006,Debabrata2013,Mondal2015,Beratan2017,GhoshWaldeck2019,Zwang2018,Torres2020}. This phenomenon has  been called Chirality-induced Spin Selectivity (CISS), and it has triggered a large amount of research in physics, chemistry, and biology, also in view of the broad spectrum of  
potential applications it may offer~\cite{ACSNano2022,Shi2004,Sun2014,Dor2013,Bustami2020,Bustami2022,Dor2017,Shinto2014,Varade2018,Chiesa2023,Santos2018}. 

There is agreement that SOC must play a key role in determining the CISS 
effect~\cite{PhysRevB.85.081404,https://doi.org/10.1002/adma.202106629,10.1063/1.3167404,Arraga2015,doi:10.1021/jp401705x,PhysRevLett.108.218102,doi:10.1021/acs.jpca.0c04562,10.1063/5.0005181,Geyer2019,karen2019} as well as time-reversal symmetry breaking via an applied voltage in transport junctions or by environmental decoherence.~\cite{PhysRevLett.108.218102,PhysRevB.93.075407,10.21468/SciPostPhysCore.6.2.044,doi:10.1021/acsnano.2c11410,PhysRevB.99.024418,doi:10.1021/acs.nanolett.0c02417,PhysRevB.107.045404} However, while time-reversal symmetry breaking is relatively easy to account for, in view of the way experiments  are performed, the origin of a  non-negligible effective SOC has not yet been fully elucidated. ~\cite{Fransson2023,Subotnik2021,Volosniev2021,Peralta2020,Peralta2023,JFransson2019,YangCaspar2020,KHuisman2023,doi:10.1021/acs.jpcc.3c08223,Dubi2022,Geyer2019} Meanwhile, the coupling to vibrational  degrees of freedom, leading to electron-vibration and spin-vibration coupling, has also been discussed.~\cite{Peralta2020,Peralta2023,PhysRevResearch.5.L022039,JFransson2019,FranssonTOptical2022}

On the side of first-principle calculations, there is no full agreement concerning the orders of magnitude of the spin polarization,~\cite{doi:10.1021/acs.jpclett.8b02360,doi:10.1021/acs.jctc.0c00621,doi:10.1021/acs.jpclett.2c03747,doi:10.1021/acsnano.2c11410,doi:10.1021/acs.jpclett.3c01922,behera2024} so that the ultimate origin of the CISS effect remains under debate. A  computational study,~\cite{behera2024} based on a fully relativistic density functional theory method combined with the Landauer-B\"uttiker approach, has suggested the need to include geometric terms in the SOC to achieve closer agreement with experimental trends. \corr{Also, a recent investigation based on a time-dependent relativistic four-current framework suggests that CISS might be related to relativistic curvature-induced helical currents and the associated magnetic fields, equally pointing to the relevance of geometrical effects.\cite{zheng2025}}

It is therefore very appealing to see  spin-orbit coupling  emerging from a simple geometric principle. In Refs. ~\cite{Shitade_2020,doi:10.1021/acs.jpclett.0c02589,doi:10.1021/acs.jpclett.4c01597} such a  geometric SOC was derived. Shitade and Minamitani~\cite{Shitade_2020} started from the Dirac Lagrangian density in a curved space-time  to arrive at an SOC expression proportional to the curvature of a helix. This SOC includes the product of the projection of the Pauli spin vector $\vec{\sigma}$ in the direction of the helix binormal vector $\vec{B}$ (using a Frenet-Serret basis), and the linear momentum $p_s$ of the electron  along the helix: $(\vec{\sigma}\cdot \vec{B})p_s$.  An estimate of the coupling strength was obtained to be approximately 160 meV, which is far stronger than any atomic SOC of light atoms. {However, there seems to be a potential issue with the approach of Shitade and Minamitani, which the authors also acknowledge: the final results can be different according to whether the thin-layer quantization is performed before or after the Foldy-Wouthuysen transformation to obtain the non-relativistic limit of the Dirac equation.}  
In fact, Yu~\cite{doi:10.1021/acs.jpclett.0c02589,doi:10.1021/acs.jpclett.4c01597}  exploited the relativistic equivalence of a curved space-time manifold and a noninertial  system to obtain a different result, in terms of the local normal vector $\vec{N}$: $(\vec{N}\times \vec{p})\cdot \vec{\sigma}$. An estimate of the coupling constant yielded in this case 0.2 meV, but for a reference polymer system with a much larger radius and pitch than, e.g., DNA. 
\corr{We remark that, although the above difficulty does not appear when the Pauli equation~\cite{PhysRevA.97.042108,doi:10.1143/JPSJ.80.073602} or the spin-independent Schr\"odinger equation~\cite{PhysRevA.23.1982,Maraner_1995} are taken as starting point, this does not solve the original problem, which so far has remained a not fully resolved mathematical issue.}

\begin{figure}[t!]
\includegraphics[width=0.99\textwidth]{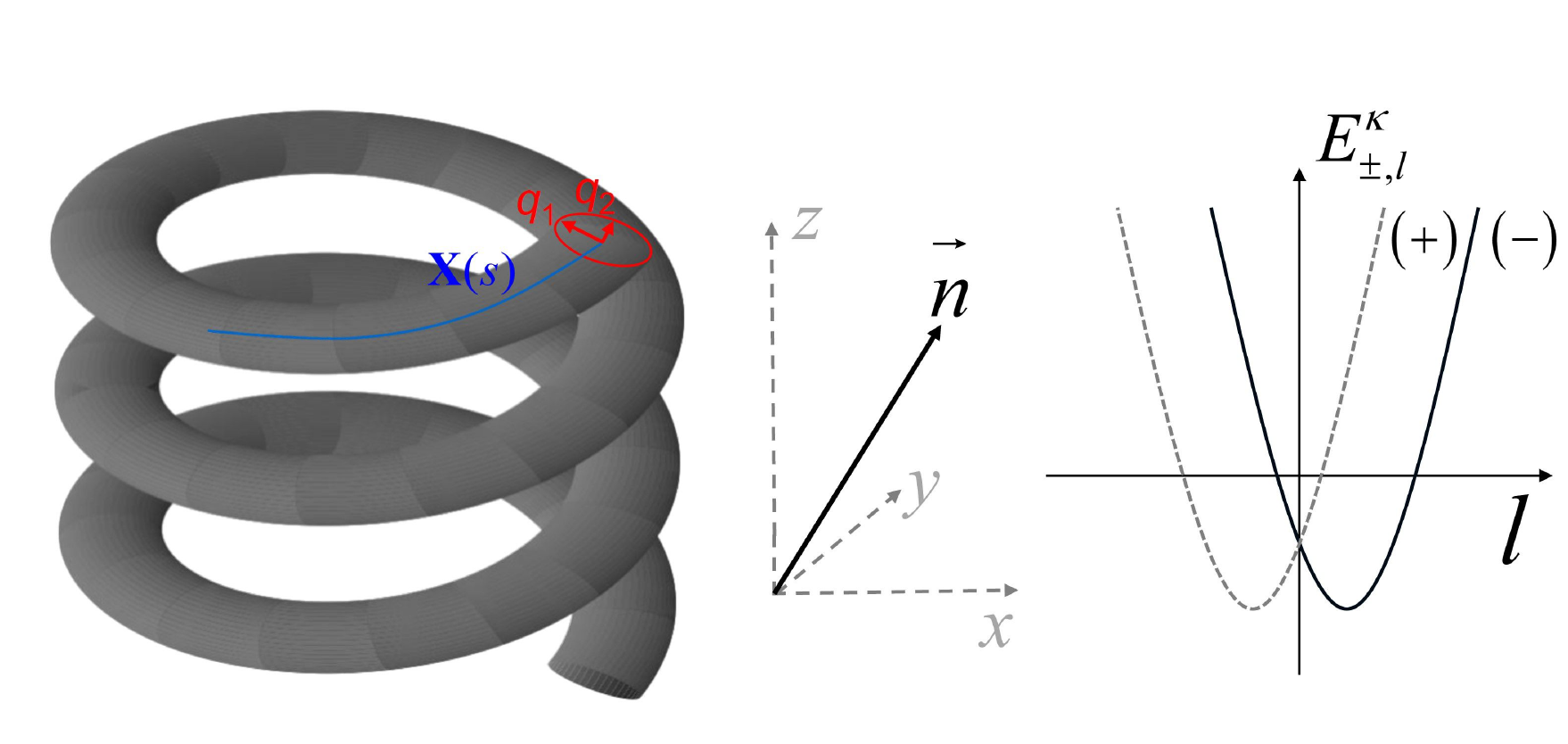}
    \caption{Schematic representation of the model system. Left panel: A helical tube is considered, which is subsequently mapped onto a one dimensional helical path following the procedure described in the text. The tube can be parametrized by the helical path described by the vector $\textbf{X}(s)$ with $s$ being the arc length, and a pair of local transversal coordinates $q_1,q_2$. In the transverse direction (cross section), a generic confinement potential $V_{\lambda}({q_1},{q_2})$ is assumed, where $\lambda$ is a measure of the strength of the confinement, similar to a quantum well. Although the specific form of $V_{\lambda}({q_1},{q_2})$ can be arbitrary, an SO(2) symmetric potential is assumed. Right panel: The eigenvalues of the Hamiltonian in Eq.~(\ref{ham1}) are schematically shown as a function of the angular momentum quantum number $l$ for the two spin branches (``+" and ``$-$"), cf. Eq.~(\ref{ev}).
    }
    \label{Fig:HelixSystem}
\end{figure}

Here we show, using a spin-independent Hamiltonian as a starting point, that the dynamics of a spinful particle along a helical path naturally develops a purely kinetic effective SOC, even if the particle does not experience any other potential (besides a spin-independent confinement potential transverse to a helical path). We find that this {chirality-induced SOC}, which we denote as $\chi$-SOC, is substantial for systems that currently show the CISS effect. We suggest that, together with the breaking of time-reversal symmetry (originating, e.g., from the external bias applied in the experiments), this $\chi$-SOC provides a simple way to generate a geometric SOC in chiral systems.
 
\section{METHODOLOGY: EFFECTIVE 1D-HAMILTONIAN ON A  HELIX}
\label{model}
We consider the Hamiltonian of an electron on a curved path, in particular, on an infinite helical tube with finite cross section (see Fig.~\ref{Fig:HelixSystem}).  
The Hamiltonian consists of a kinetic energy term $\hat T$ and a transverse confinement potential $V$, but no spin-dependent interactions are considered. 
\corr{When applied to real molecular systems, the confinement  potential is related to the electrostatic potential distribution along the molecular frame, and it is, thus, dependent on the chemical composition. However, at the level of abstraction we are working, our choice is guided by Occam's razor, so that we  impose few minimal conditions: the confinement potential should (i) be a continuous function, (ii) have a minimum on all points along the helical pathway shown in Fig~\ref{Fig:HelixSystem}, (iii) allow for an analytically closed solution of the transverse Schr\"odinger equation, i.e., it should not depend on the arc length $s$, and (iv) be spin-independent.} 

\corr{ Following a procedure presented originally by da Costa,~\cite{PhysRevA.23.1982}, but also formalized by e.g., Maraner,~\cite{Maraner_1995} and others (see e.g. \cite{Kimouche2024,OUYANG1999297}), one can decouple longitudinal (along the helical path) and transverse degrees of freedom to map the 3d-Hamiltonian structure of the helical tube on an effective 1d-Hamiltonian. This procedure has, more recently, been formalized by Geyer et al.~\cite{10.1063/5.0005181} within a rigorous space-adiabatic framework, \corr{$-$a similar approach has also been used, e.g., in Refs.~\cite{karen2019,PhysRevA.101.053632,PhysRevA.97.042108}.} We skip here the  details of the derivation, which are provided in the Supplementary Information (SI) section for three different confinement potentials: a)  an SO(2) harmonic confinement, b) square well harmonic confinement, and c) a square well hard wall confinement. Here, we limit ourselves to present the results for the SO(2) symmetric potential. }

The obtained effective 1d-Hamiltonian reads: 

\begin{equation}\label{Ham}
\hat{H}^{1d}_{eff}=-\frac{\hbar^2}{2mL^2} \{ \frac{\partial^2}{\partial\phi^2} +\frac{\rho R}{4} \},
\end{equation}

\noindent where $\rho=R/L^2$ and $\tau=(b/2\pi)/L^2$ are the curvature and torsion, respectively, of a helix with radius $R$ and pitch $b$. $L=\sqrt{R^2+(b/2\pi)^2}$ is related to the length of a single helical turn $L_0$ via $L=L_0/2\pi$.  
 The second term in Eq.~(\ref{Ham}) is a quantum geometric potential already obtained by da Costa,~\cite{PhysRevA.23.1982} but also in other studies.~\cite{10.1063/5.0005181,karen2019,PhysRevA.101.053632,PhysRevA.97.042108,Maraner_1995} Additional terms proportional to the torsion of the path may also appear for other choices of the confinement potential (see the SI section). However, they are all spin diagonal.

Consider now the general representation of a spinor wave function on the 1d helical path~\cite{note4} and with a Hamiltonian described by Eq.~(\ref{Ham}): 
\begin{equation}\label{wf}
\vec{\Psi}(\phi)= \exp{-i \kappa\frac{\phi}{2} \vec{n}\cdot\vec{\sigma}}\vec{\chi}\otimes\Phi(\phi) =\mathcal{U}(\phi,\vec{n}) \vec{\chi}\otimes\Phi(\phi) .
\end{equation}
 Due to the absence of SOC in Eq.~(\ref{Ham}) the spin and spatial components are separable. The 2-component spinor $\vec{\chi}$ does not need to be specified at this stage, its components will be calculated later on. Notice that the spin rotation is tied to the space frame of the helix, which is also a result of the confinement potential used to enforce the electron to follow the helical pathway. \corr{In other words, the very presence of the helix constrains the 
 electron to follow the curved path, with the spin following suit.} The unitary operator acting on the  spinor $\vec{\chi}$  induces a spin rotation around $\vec{n}$ while the electron moves along the helix (this is similar to the action of a quantum gate on a qubit, with the helix playing the role of the ``quantum gate''). The parameter $\kappa=\pm 1$ accounts for a change from a right-handed to a left-handed helix, since the sign of $\phi$ changes in this case.  Notice that the quantum geometric potential commutes with the spin rotation operator, and hence it will have no influence on the spin-dependent properties of the model.
 
 The spatial part $\Phi(\phi)$ can be written as a linear combination of ``plane wave'' solutions with (real-valued) angular momentum $l$ as: 
 \begin{equation}\label{basis}
 \Phi(\phi)=\int_{-\infty}^{\infty} dl \,A_l e^{il\phi}.
 \end{equation}
 Here, the normalization condition is $\int_{0}^{2\pi} (d\phi/2\pi) |\Phi(\phi)|^2=1$, from which it follows that $\int_{-\infty}^{\infty} dl \,|A_l|^2 =1$.
For the calculations of the charge and spin currents later on, it is convenient to restrict the integration to positive values of $l$ by introducing the index $s=\mathrm{sgn}(l)=\pm 1$: $e^{il\phi}\rightarrow e^{is|l|\phi}$. 
Acting with the Hamiltonian Eq.~(\ref{Ham}) on the wave function Eq.~(\ref{wf}), and defining $E_0 =\hbar^2/2mL^2$ as a characteristic energy scale of the problem, we obtain the following:

\begin{eqnarray}
{\hat{H}^{1d}_{eff}}\vec{\Psi}(\phi)&=&
E_0 \mathcal{U}(\phi,\vec{n})  \{ (-i\frac{\partial}{\partial\phi})^2- \kappa (\vec{n}\cdot\vec{\sigma}) (-i\frac{\partial}{\partial\phi}) \\ \nonumber 
&+&\frac{1}{4}(\vec{n}\cdot\vec{\sigma})^2 - \frac{\rho R}{4}\} \vec{\chi}\otimes\Phi(\phi) \\ \nonumber
&=& E_0 \mathcal{U}(\phi,\vec{n}) \{ p_{\phi}^2  - \kappa (\vec{n}\cdot\vec{\sigma}) p_{\phi} \\ \nonumber 
&+&(\frac{1}{4} - \frac{\rho R}{4}) \} \vec{\chi} \otimes \Phi(\phi).
 \end{eqnarray}

In the second equality, we have introduced the angular momentum operator $p_{\phi}=-i\partial/\partial\phi$ and used the result $(\vec{n}\cdot\vec{\sigma})^2=1$. Therefore, we can introduce 
 a new Hamiltonian as:
\begin{eqnarray}
\hat{H}_{1d}=E_0 \{ [p_{\phi}^2 +\frac{1}{4}(1 - \rho R)] \mathds{1}_{2\times 2} - \kappa (\vec{n}\cdot\vec{\sigma}) p_{\phi}   \},
\label{ham1}
 \end{eqnarray}
 where we stress the fact that the kinetic energy operator and the correction leading to a geometric potential are both  diagonal in spin space. These results show that it is  possible to derive an effective spin-orbit coupling for an electron moving on a curvilinear path,{ even if no previous SOC was present}. The key result is that the geometric phase accumulated by the spin during its motion leads to an effective interaction between the spin and the orbital degrees of freedom, which can be interpreted as a chirality-induced spin-orbit coupling term: 
 \begin{equation}
 {(\vec{L}}\cdot {\vec{S})}_{\mathrm{\chi-SOC}}\equiv(2E_0/\hbar)\kappa (\vec{n}\cdot\vec{S}) p_{\phi}. 
 \end{equation}
 
 Notice that the obtained $\chi$-SOC has a purely kinetic origin and its strength is controlled by the energy scale $E_0$. For a DNA helix with $R=1 \, nm$ and $b=3.4 \, nm$, one estimates $E_0\approx 30\, meV$, which is larger by a factor 3 to 4 than the atomic SOC of light elements.\footnote{However, note that in this estimate we have used the free electron mass $m$. In a more accurate calculation, an effective mass $m^{*}$ should be employed.} 
 The obtained geometric SOC is clearly time-reversal invariant, and it can also be rewritten as an SU(2) {``pseudo-gauge field''} by completing squares in Eq.~(\ref{ham1}): 
\begin{eqnarray}
\hat{H}_{1d}&=&E_0 (p_{\phi} \mathds{1}_{2\times 2} - e{A}_{SO})^{2} - \frac{\hbar^2}{8m}\rho^2,
\end{eqnarray}
with ${A}_{SO}=\frac{\kappa}{2e} \vec{n}\cdot\vec{\sigma}$. This approach leverages the properties of SU(2) rotations to capture the evolution of the spin state in a curved trajectory. Note that this contribution would also be present on a circle ($b=0$), although in this case the angular momentum variable would be quantized due to the periodicity condition $\Phi(\phi+2\pi)=\Phi(\phi)$, but  it would trivially vanish as $R\rightarrow\infty$, i.e., in the limit of a straight line. 

\section{ RESULTS AND DISCUSSION}
\label{result}

The eigenvalues of Eq.~(\ref{ham1}) can be easily found: 
\begin{eqnarray} 
\tilde{E}^{\kappa}_{\pm,l} = l^2 + \frac{1}{4}(1-\rho R) \pm \kappa l 
= (l \pm \frac{\kappa}{2})^2 -\frac{\rho R}{4}, 
\label{ev}
 \end{eqnarray}

\noindent where $\tilde{E}^{\kappa}_{\pm,l}=(E^{\kappa}_{\pm,l})/E_0$. This represents two parabolas shifted horizontally from each other by $\kappa$ (see schematic in Fig.~\ref{Fig:HelixSystem}). The eigenvalues as a function of $l$ yield two spin branches (``+" and ``$-$") for positive $l$ (or $s$ = 1), and another two spin branches for negative $l$. Due to time-reversal symmetry, the relation $\tilde{E}^{\kappa}_{+,l}=\tilde{E}^{\kappa}_{-,-l}$ is valid, so that Kramer's theorem holds, as expected.

The corresponding spinor eigenfunctions $\vec{\chi}$ can be obtained in terms of the components of the  vector $\vec{n}$, which we parametrize in general using two angles $\alpha,\beta$ as $\vec{n}=(\sin\alpha\cos{\beta}, \sin\alpha\sin{\beta}, \cos\alpha)$. In this way, we get the following: 
\begin{eqnarray}
\vec{\chi}_{+}=e^{i\beta/2}\left(
\begin{array}{l}
\sin{(\alpha/2)}e^{-i\beta/2}\\
-\cos{(\alpha/2)}e^{i\beta/2}
\end{array}
\right),
\label{ew1}
 \end{eqnarray}

 and 
\begin{eqnarray}
\vec{\chi}_{-}=e^{-i\beta/2}\left(
\begin{array}{l}
\cos{(\alpha/2)}e^{-i\beta/2}\\
\sin{(\alpha/2)}e^{i\beta/2}
\end{array}
\right).
\label{ew2}
 \end{eqnarray}

Another pair of eigenvectors is obtained for $s=-1$  simply by replacing $e^{is|l|\phi}\rightarrow e^{-is|l|\phi}$.

We can now calculate, in the local $\phi$-frame,  the average charge current $j^{\pm}_{c}$ and the average spin current $j_{spin}$:
\begin{eqnarray}
  j^{\pm}_{c}&=&{\int_{0}^{2\pi} \frac{d\phi}{2\pi}\,} \vec{\Xi}_{\pm}^{\dagger}(\phi) (e\hat{\nu_{\phi}}/L)\vec{\Xi}_{\pm}(\phi), \\
  j_{spin}&=&(1/4)\hbar\sum_{j=\pm} {\int_{0}^{2\pi} \frac{d\phi}{2\pi}}\, \vec{\Xi}_{j}^{\dagger}(\phi) \{\hat{\nu}_{\phi},\sigma_{z}\}\vec{\Xi}_{j}(\phi),
\end{eqnarray}

with $\{ \dots\}$ being the anticommutator.~\cite{Transport-book} 
Here, we have defined $\vec{\Xi}_{\pm}(\phi)=\vec{\chi}_{\pm}\Phi(\phi)$. \corr{The velocity operator $\hat{\nu}_{\phi}=(i/\hbar)[\phi,\hat H_{1d}]$,} defined on the basis of Eq.~(\ref{ham1}), contains a spin-dependent part, and it is given by: 
\begin{eqnarray}
\hat{\nu}_{\phi}=\frac{E_0 L}{\hbar}\{ -2i\frac{d}{d\phi} \mathds{1}_{2\times 2} - (s\kappa)\, \vec{n}\cdot\vec{\sigma} \}.
\label{veloc}
 \end{eqnarray}

Using the latter expression, the charge current is: 
\begin{eqnarray}
j^{\pm,\kappa,s}_{c}&=&\frac{e\hbar}{mL^2} s(\bar{l} \pm \frac{\kappa}{2}),
\end{eqnarray}
with $\bar{l}=\int_{0}^{\infty} dl \,|A_l|^2 l$. This gives a total charge current of $j^{\kappa,s}_{c}=(2e\hbar/mL^2) s \bar{l}$.~\footnote{Notice that for systems with a very small pitch, transport pathways along the helical axis may interfere (due to tunneling) with those following a helical path, and eventually lead to a reduction of chirality-induced effects. This is, indeed the case of helicene, as shown in Ref.~\cite{doi:10.1021/acs.jpclett.8b00208} using first-principles calculations.}

The difference of these currents for a given propagation direction, e.g., $s=1$, yields $(e\hbar/mL^2) s\kappa$, which is proportional to the helicity $\kappa$, and thus changes sign upon a mirror inversion operation. The fact that this difference does not vanish indicates that the spins in the (+) and ($-$) states propagate with different velocities. This, in particular, allows us to define a spin polarization ($\mathrm{SP}$) of the charge current: 

\begin{equation}
    \mathrm{SP}=(j^{+,\kappa,s}_{c}-j^{-,\kappa,s}_{c})/j^{\kappa,s}_{c}=\kappa/2\bar{l}.
    \label{sp}
\end{equation}
We remark that in a real system, finite-size quantization will lead to discrete values of the $l$ quantum number and the integrals will become summations: $\int_{0}^{\infty} dl\rightarrow\sum_{l=1}^{\infty}$. The representation of the wave function  $\Phi(\phi)$ will then read: $\Phi(\phi)=(1/\sqrt{C})\sum_{l=-\infty}^{\infty}A_le^{il\phi}$, with $C=\sum_{l=-\infty}^{\infty}|A_l|^2$ to ensure proper normalization.

In the special case of a single $l$-mode contributing to the summation, we can make a rough estimate of the spin polarization by assuming $|A_l|^2\sim\delta(l-l_0)$, so that $\mathrm{SP}=\kappa/(2l_0)$, which for $l_0=1$ gives a  50\% polarization for $\kappa=1$. This is of the same order of magnitude of the measured spin polarizations in, e.g., DNA~\cite{Ray1999}.

Another simple example is a Gaussian profile with standard deviation $\sigma$ with $|A_l|^2=(1/\sqrt{2\pi\sigma^2}) \; \exp{(-(l-l_0)^2/2\sigma^2 )}$, which yields: 
\begin{equation}
\bar{l}=\int_{0}^{\infty} dl \,|A_l|^2 l=
\frac{l_0}{2} \left[1 + \operatorname{erf}\left( \frac{l_0}{\sqrt{2\sigma^2}} \right)\right]
+ \frac{\sigma}{\sqrt{2\pi}} \exp\left( -\frac{l_0^2}{2\sigma^2} \right).
\end{equation}

\begin{figure}[t!]
\includegraphics[width=0.99\textwidth]{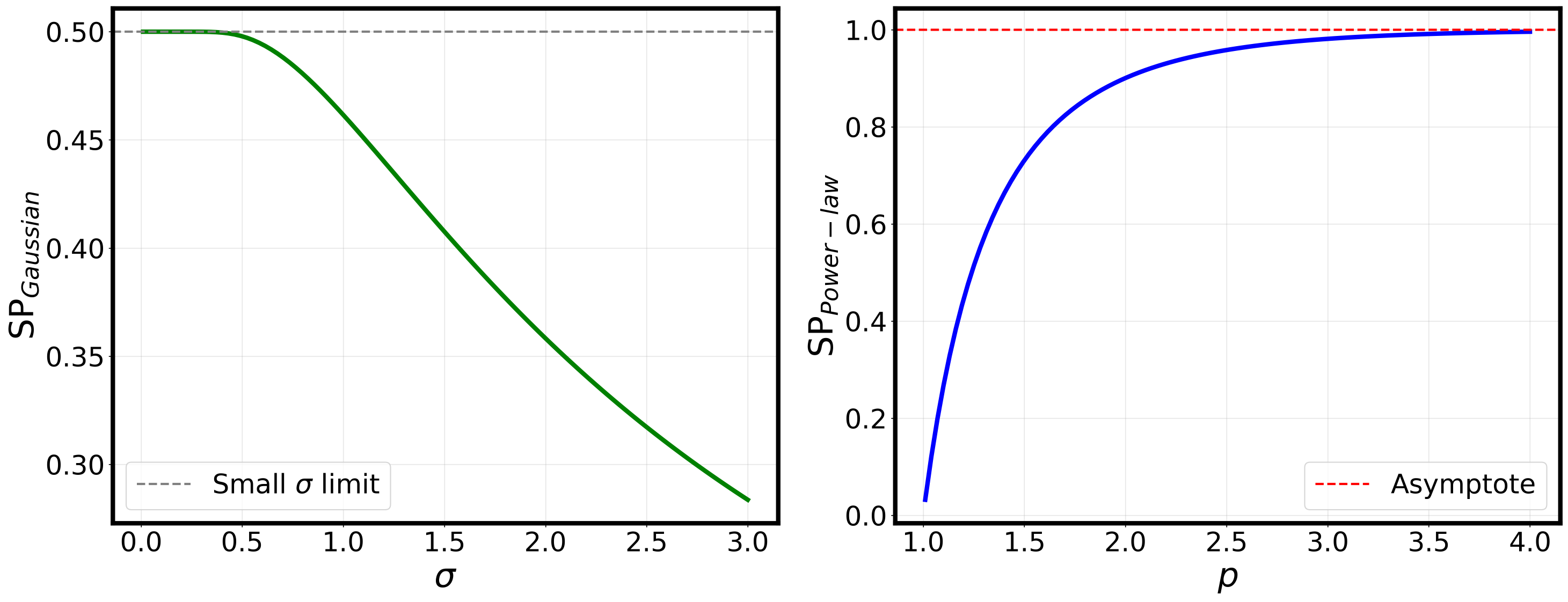}
    \caption{Spin polarization $\mathrm{SP}$ calculated according to Eq.~(\ref{sp}) for the two assumed functions of the amplitudes $A_l$. Left panel: $\mathrm{SP}$ for the Gaussian model as a function of the variance $\sigma$. Dashed line corresponds to the limit of small variance.  Right pannel: $\mathrm{SP}$ for the power-law model as a function of the exponent $p$. The dashed line corresponds to the limit of large $p$.
    }
    \label{Fig:sp}
\end{figure}

For small $\sigma$, $\bar{l}\approx l_0 + O(\sigma^3)$, so that we recover the previous result $\mathrm{SP}=\kappa/2l_0$=50\% for $l_0=1$. In the opposite case ($\sigma\gg l_0$), one gets asymptotically $\bar{l}\approx \frac{\sigma}{\sqrt{2\pi}}  + \frac{l_0}{2}$, which leads to $\mathrm{SP}\approx\kappa/(\sqrt{\frac{2}{\pi}}\sigma  + l_0)\approx \kappa \sqrt{\frac{\pi}{2}} (1/\sigma) + O(\sigma^{-3})$. In this case, the polarization can become much smaller than 50\% (the case for a single $l$-value), depending on the value of $\sigma$ (see Fig.~\ref{Fig:sp}, left panel).

As a last example, more appropriate for a discrete $l$-set, consider a power-law decay of the coefficients $A_l\propto l^{-p}$. The resulting summations can be obatined in closed form in terms of Riemann´s $\zeta(p)$-function. One gets  $\mathrm{SP}=\kappa\zeta(2p)/2\zeta(2p-1)$, with $p>1$. This ratio yields already SP$\approx 77\%$ for $p=1.5$, and then rapidly converges to 100\% for larger $p$-values (see Fig.~\ref{Fig:sp}, right panel).

Notice, however, that in all three examples we have assumed broken time-reversal symmetry, since there would always be another mode with negative $l$ yielding the opposite polarization. This can be more clearly seen from the shape of the dispersion relation, see Fig~\ref{Fig:HelixSystem}, and is a consequence of Kramer's theorem.

 More realistic estimates would require the formulation of a full spin transport problem with inclusion of scattering effects at, e.g., substrate-molecule interfaces, and those due to interactions with, e.g., vibrational degrees of freedom.~\cite{PhysRevResearch.5.L022039,VarelaScipost2023} {We also point out that even though the magnitude of spin polarization may be different, the phenomenon we predict is present irrespective of whether the transport mechanism is ballistic or hopping~\cite{Transport-book}. The reason is that even in the latter case the helix would create an effective spin-orbit coupling of the type we derive.}

In a similar way,  the spin currents can be calculated, yielding: 
\begin{eqnarray}
j^{\pm,\kappa,s}_{spin}(\alpha)&=&\mp\frac{\hbar^2}{2mL} s\cos{\alpha}\, (\bar{l}\pm\frac{\kappa}{2}),
\end{eqnarray}
which lead  to the total spin current  
$j^{\kappa,s}_{spin}(\alpha)=-(\hbar^2/2mL)\kappa s\cos{\alpha}$.  Using the helix parameters above of DNA, we can obtain an estimate of the spin current coefficient $\hbar^2/2mL=0.332\,eV\,nm$. 

Notice that the symmetries of the spin current are: $j^{\pm,-\kappa,s}_{spin}(\alpha)=-j^{\pm,\kappa,s}_{spin}(\alpha)$, and $j^{\pm,-\kappa,-s}_{spin}(\alpha)=j^{\pm,\kappa,s}_{spin}(\alpha)$, i.e., changing the chirality changes the sign of the spin current, while a change in chirality together with time-reversal ($s\to -s$) leaves the spin current invariant. A nonzero spin current only emerges if time-reversal symmetry is broken; otherwise any contribution for $+s$ will be canceled by a term similar to $-s$.

In a geometric picture, the chirality parameter $\kappa$ should be related to the helix torsion (or to the pitch) in a suitable dimensionless quantity. It is also interesting to note that our results have a qualitative resemblance to  the analytical model presented in Ref.~\cite{10.1063/1.4921310}, which, however, introduces a Rashba spin-orbit interaction in the Hamiltonian from the start. Moreover, the angle $\alpha$ in our case parameterizes the spin rotation vector, while in Ref.~\cite{10.1063/1.4921310} it is related to the strength of the Rashba spin-orbit coupling.

\section{CONCLUSIONS}
\label{conclusion}

In conclusion, we have shown that a spinful particle traveling along a helical path naturally develops an effective SOC, even without an intrinsic SOC. This chirality-induced SOC ($\chi$-SOC) is stronger than the typical relativistic SOC of light atoms, thus providing an additional source of spin polarization. Our results suggest a possible strong additional contribution to the CISS effect observed experimentally in chiral organic and inorganic materials with intrinsic helical topologies (DNA, $\alpha$-helices, helicene and its derivatives, chiral crystals, etc.). In future work, it would be interesting to address issues like the temperature and length dependence of this effect, which would require a spin transport calculation, eventually including the interaction with dynamical degrees of freedom such as linear or chiral phonons. Regardless of the effect, \corr{our work suggests that geometric effects introduced by a chiral structure can generate a spin-orbit coupling contribution, which is otherwise absent in systems where mirror symmetry is not broken.}
It is finally worth mentioning that our model Hamiltonian can be formally adapted to different spin transport setups, including  two-terminal measurements, as in break junctions,~\cite{https://doi.org/10.1002/smll.201602519} as well as  more complex setups, which address a "transverse" CISS effect.~\cite{PhysRevLett.133.108001,doi:10.1021/jacs.1c02983} We leave this for future studies.

\section*{ASSOCIATED CONTENT}
\subsection*{Supporting Information}
The Supporting Information is available free of charge at XXX. \\

Derivation of the effective 1D Hamiltonian using the confinement potential approach

\section*{AUTHOR INFORMATION}

\noindent {\bf Corresponding authors}\\
Massimiliano Di Ventra $-$ Department of Physics, University of California San Diego, La Jolla, CA, 92093, USA; https://orcid.org/0000-0001-9416-189X; Email: diventra@physics.ucsd.edu. \\
Rafael Gutierrez $-$ Institute for Materials Science and Max Bergmann Center of Biomaterials, TU Dresden,
01062 Dresden, Germany; https://orcid.org/0000-0001-8121-8041; Email: rafael.gutierrez@tu-dresden.de. \\
Gianaurelio Cuniberti $-$ Institute for Materials Science and Max Bergmann Center of Biomaterials, TU Dresden, 01062 Dresden, Germany and Dresden Center for Computational Materials Science (DCMS), TU Dresden, 01062; https://orcid.org/0000-0002-6574-7848; Email: gianaurelio.cuniberti@tu-dresden.de. \\

\noindent {\bf Author contributions}\\ $^{a}$ M. Di V. and R. G. contributed equally to this work. \\

\noindent{\bf Notes} \\
The authors declare no competing financial interest.

\section*{ACKNOWLEDGMENTS}
The authors thank Matthias  Geyer for very fruitful discussions. M. D. acknowledges funding by the Alexander von Humboldt Stiftung through the 2024 Humboldt Research Award. R.G. and G.C.  acknowledge the support of the German Research Foundation (DFG) within the project Theoretical Studies on Chirality-Induced Spin Selectivity (CU 44/55-1), and by the transCampus Research Award Disentangling the Design Principles of Chiral-Induced Spin Selectivity (CISS) at the Molecule-Electrode Interface for Practical Spintronic Applications (Grant No. tCRA 2020-01) and Programme trans-Campus Interplay between vibrations and spin polarization in the CISS effect of helical molecules (Grant No. tC2023-03).


\begin{mcitethebibliography}{82}
\bibitem[{}()]{note4}
{} Note that due to the homomorphism between the SU(2) group of spin rotations and the SO(3) group of spatial rotations, the angle $\phi$ in the unitary operator $\mathcal{U}(\phi)$ is the same as the one parametrizing the arc length along the helix $s=\phi/L$. This may be intuitively understood by considering the mapping $\mathcal{U}(\phi,\vec{n})(\vec{u}\cdot\vec{\sigma}) \mathcal{U}^{\dagger}(\phi,\vec{n}) = (R(\phi,\vec{n})\vec{u})\cdot \vec{\sigma} $. Here, the SO(2) unitary operator $\mathcal{U}(\phi,\vec{n})$ has been defined in Eq.~\ref{wf}, $\vec{u}$ is an arbitrary vector in 3d space, and $R(\phi,\vec{n})$ is an element of the non-Abelian SO(3) group.\relax
\mciteBstWouldAddEndPunctfalse
\mciteSetBstMidEndSepPunct{\mcitedefaultmidpunct}
{}{\mcitedefaultseppunct}\relax
\EndOfBibitem
\end{mcitethebibliography}
\providecommand{\latin}[1]{#1}
\makeatletter
\providecommand{\doi}
  {\begingroup\let\do\@makeother\dospecials
  \catcode`\{=1 \catcode`\}=2 \doi@aux}
\providecommand{\doi@aux}[1]{\endgroup\texttt{#1}}
\makeatother
\providecommand*\mcitethebibliography{\thebibliography}
\csname @ifundefined\endcsname{endmcitethebibliography}  {\let\endmcitethebibliography\endthebibliography}{}


\end{document}


\title{Supplementary Information \ \\ \ \\ Chirality-induced Spin-Orbit Coupling and Spin Selectivity}

\author{Massimiliano Di Ventra}
\email{diventra@physics.ucsd.edu}
\affiliation{Department of Physics, University of California San Diego, La Jolla, CA, 92093, USA}
\author{Rafael Gutierrez}
\email{rafael.gutierrez@tu-dresden.de}
\affiliation{Institute for Materials Science and Max Bergmann Center of Biomaterials, TU Dresden, 01062 Dresden, Germany}
\author{Gianaurelio Cuniberti}
\email{gianaurelio.cuniberti@tu-dresden.de}
\affiliation{Institute for Materials Science and Max Bergmann Center of Biomaterials, TU Dresden, 01062 Dresden, Germany}
\affiliation{Dresden Center for Computational Materials Science (DCMS), TU Dresden, 01062 Dresden, Germany}
\clearpage

\maketitle

{
\noindent \textbf{Derivation of a 1-dimensional Hamiltonian on a helical pathway} \\
\label{hamiltonian}

Since we are only considering kinetic energy terms in the initial (spin-diagonal) Hamiltonian, our starting point is the general form of the Laplacian operator in general curvilinear coordinates (in units of ${\hbar^2}/(2m)$):
\begin{equation}
  \hat T = \frac{1}{{\sqrt g }}{\partial _n}(\sqrt g {G^{nm}}{\partial _m})
  \label{eq1}
\end{equation}

Here, $g$ is the (positive) determinant of the associated metric tensor and $G$ its inverse. For the helical tube shown in Fig. 1 of the main text, we can choose a vector basis parametrized with the arc length $s$. Any neighborhood of a helical path described by a vector \textbf{X}(s) can thus be generally written as: $\textbf{R}(s) = \textbf{X}(s) + {q_1}\textbf{N}(s) + {q_2}\textbf{B}(s)$ in terms of the unit normal vector \textbf{N}(s) and binormal vector \textbf{B}(s) to the helical path in a Frenet-Serret frame. The metric tensor can be obtained as ${g_{nm}} = {\partial _n}\textbf{R} \cdot {\partial _m}\textbf{R}$ with $n=s,q_1,q_2$, resulting in:

\begin{equation}
    g_{\epsilon} = \left( {\begin{array}{*{20}{c}}
{{A_{\epsilon}^2} + \epsilon{\tau ^2}(q_1^2 + q_2^2)}&{ - {\epsilon}\tau {q_2}}&{\epsilon\tau {q_1}}\\
{ - \epsilon\tau {q_2}}&\epsilon&0\\
{\epsilon\tau {q_1}}&0&\epsilon
\end{array}} \right),
\end{equation}

with $A_{\epsilon}=\sqrt{f}=({1-\sqrt{\epsilon}\rho q_1})^{1/2}$. Notice that this basis is non-orthogonal. The corresponding inverse $G_{\epsilon}$ is given by:

\begin{equation}
    G_{\epsilon} = \frac{1}{A_{\epsilon}^4}\left( {\begin{array}{*{20}{c}}
1 &  { \tau {q_2}} &  {-\tau {q_1}}\\
{ \tau {q_2}}  &  \frac{1}{\epsilon}[{{{A_{\epsilon}}^2} + {\tau ^2}q_2^2 }]&  {-\tau ^2}q_1q_2  \\
{-\tau {q_1}}   &   {-\tau ^2}q_1q_2     & \frac{1}{\epsilon}[{{{A_{\epsilon}}^2} + {\tau ^2}q_1^2 }]
\end{array}} \right).
\end{equation}

For convenience (see the subsequent developments below), we have already rescaled the transverse coordinates $q_1,q_2$ with a small factor $\sqrt{\epsilon}$, $q_{1,2}\rightarrow \sqrt{\epsilon} q_{1,2}$ which will control the strength of the transverse confinement potential and provide a natural expansion parameter.

Our aim is to obtain an effective 1-dimensional Hamiltonian by projecting out transversal degrees of freedom. For this, a confinement potential ${V_\lambda }({q_1},{q_2})$ is added to the kinetic energy operator $\hat T$, with $\lambda$ being a measure of the confinement, and the (formal) limit $\lambda\to\infty$ considered. 

The specific form of the confinement potential, is in general terms, arbitrary. When applied to real molecular systems, this potential is related to the electrostatic potential distribution in the molecular frame, and it is, thus, dependent on the chemical composition. However, at the level of abstraction we are working, our choice is guided by Occam's razor, so that we  assume few minimal conditions: the confinement potential should (i) be a continuous function, (ii) have a minimum on all points along the helical pathway, (iii) allow for an analytically closed solution of the transverse Schr\"odinger equation to be defined later, i.e., it does not depend on the arc length $s$, and (iv) be spin-independent. 

Then, to lowest order in a Taylor expansion, similar to what is done for modelling quantum wells, we chose a harmonic confinement with SO(2) rotational symmetry (another example is briefly discussed at the end of this section): ${V_\lambda }({q_1},{q_2}) = {\lambda ^2}(q_1^2 + q_2^2)/2=\lambda^2r^2/2= \lambda ^2 r^2/2\varepsilon$, where in the last step the coordinate rescaling has been introduced. Now, $\lambda$ can remain finite, while the strength of the confinement is absorbed in $\epsilon$. 

For small enough, but finite, $\epsilon$ (strong confinement) the system will remain in its transverse ground state and hence, only the ground state wave functions $\Phi_0(q_1,q_2)$ of the transverse Schr\"odinger equation will need to be considered. As shown in Refs. [1,4,5], a mathematically clean procedure to project the full Hamiltonian on the lowest transverse states, thus leading to an effective one-dimensional problem, can be defined as: 

\begin{equation}
{\hat H_{eff}} = \mathop {\lim }\limits_{\varepsilon  \to 0} \left\langle {{\Phi _{0}}({q_1},{q_2})| A_\varepsilon ^{1/2} \hat H A_\varepsilon ^{ - 1/2} - {\hat H_ \bot }{\rm{|}}{\Phi _{0}}({q_1},{q_2})} \right\rangle
\end{equation}

The term H$_{\perp}$ is the part of the Hamiltonian including only transverse degrees of freedom $-$scaling as $\epsilon^{-1}$$-$ and which will be defined below. This procedure is more systematic than the approach used by Da Costa [2], although it leads to similar results for an infinitely strong confinement ($\epsilon\to 0$). 

Based on these preliminaries, we can now use Eq. (\ref{eq1}) to determine $A_\varepsilon ^{1/2} \hat H A_\varepsilon ^{ - 1/2}$. Expanding it, we obtain: 

\begin{eqnarray}
\hat H_{eff}&=&A_\varepsilon ^{1/2} \hat H A_\varepsilon ^{ - 1/2}=\frac{1}{A_\varepsilon}\partial_n[({A_\varepsilon^2 G^{nm} \partial_m})\frac{1}{A_\varepsilon}]\\
&=& G^{nm} \{\frac{1}{A_\varepsilon} (\partial_n A_\varepsilon^2)(\partial_m \frac{1}{A_\varepsilon}) + A_\varepsilon (\partial_n \partial_m \frac{1}{A_\varepsilon}) \}  \\
&+&A_\varepsilon (\partial_n G^{nm}) (\partial_m \frac{1}{A_\varepsilon}) + \{\frac{1}{A_\varepsilon}(\partial_n A_\varepsilon) G^{nm} + (\partial_n G^{nm})  \} \partial_m \nonumber\\
&+& A_\varepsilon G^{nm} (\partial_m \frac{1}{A_\varepsilon}) \partial_n + G^{nm} \partial_n \partial_m + V_\epsilon(q_1,q_2). \nonumber
\label{metric}
\end{eqnarray}

A very lengthy, but straightforward calculation yields:
\begin{eqnarray}
\hat H_{eff}&=&\frac{1}{A_\varepsilon^4}\{ \partial_s -\tau (q_1\partial_2-q_2\partial_1)\}^2 -\frac{1}{\epsilon}\frac{\partial^2_1 A_\varepsilon}{A_\varepsilon} \nonumber \\
&&  - 4\frac{\partial_1 A_\varepsilon}{A_\varepsilon} \frac{\tau}{A^4_\varepsilon}q_2\{ \partial_s-\tau (q_1\partial_2 - q_2\partial_1) \} \nonumber \\
&& + \frac{\tau^2}{A_\varepsilon^4} \{ [ (\frac{\partial_1 A_\varepsilon}{A_\varepsilon})^2 -\frac{\partial^2_1 A_\varepsilon}{A_\varepsilon}  ]q^2_2 + \frac{\partial_1 A_\varepsilon}{A_\varepsilon} q_1 \}  + \frac{1}{\epsilon}(\partial_1^2+\partial_2^2)  + V_\epsilon (q_1,q_2) \nonumber \\
&&= \hat T + \frac{1}{\epsilon}(\partial_1^2+\partial_2^2)  + V_\epsilon(q_1,q_2),
\label{metric1}
\end{eqnarray}

from where we can identify ${H_ \bot }$ as:
\begin{eqnarray}
{\hat H_ \bot }= \frac{1}{\epsilon}(\partial_1^2+\partial_2^2) + V_\epsilon (q_1,q_2). 
\label{trans}
\end{eqnarray}

In the next step, we will perform an expansion in $\epsilon$ before proceeding to build matrix elements with the transverse wave functions, which can be obtained analytically by solving Eq. (\ref{trans}). For the expansion, we will use the result: 

\begin{equation}
  \frac{1}{(1-\sqrt{\epsilon}C )^p}  \approx 1- p \sqrt{\epsilon}C + \frac{p(p+1)}{2} \epsilon C^2 - \frac{p(p+1)(p+2)}{6} \epsilon^{3/2} C^3 + O(\epsilon^2), 
\end{equation}

and introduce the transverse angular momentum operator $\hat L=-i(q_1\partial_2-q_2\partial_1)$.  Applying this procedure, we first expand the prefactors $(A_\epsilon)^p=f_\epsilon^{p/2}$ in powers of  $\epsilon$ up to order $\epsilon^{1/2}$ (we assume $\epsilon$ to be small enough, so that  cutting the expansion at this order is a valid approximation):

\begin{eqnarray}\label{teps}
 \hat T_\epsilon &=& [1-2  \sqrt{\epsilon}\rho q_1 ](\{ \partial_s - i\tau \hat L\}^2 \\
 &&+ \frac{\rho^2}{4}[1-\frac{3}{2}  \sqrt{\epsilon}\rho q_1]] \nonumber \\
 && + \sqrt{\epsilon} \{ 4\rho\tau [[1-\frac{5}{2}  \sqrt{\epsilon}\rho q_1]  q_2(\partial_s-i\tau \hat L) - \tau^2\rho q_1 [1-3  \sqrt{\epsilon}\rho q_1] \} \nonumber \\
 && + \epsilon \frac{\tau^2\rho^2}{2} q_2^2 [1-4  \sqrt{\epsilon}\rho q_1]
 + O(\epsilon). \nonumber
\end{eqnarray}

In the next step, we need to build matrix elements of Eq.~(\ref{teps}) over the ground state wave functions of the transverse Hamiltonian ${H_ \bot }$, which, for the assumed SO(2) potential, are given by 

\begin{eqnarray}
    |\Phi_{l,0}(\theta)>=N_l e^{il\theta} (\lambda r)^{|l|}e^{-\lambda^2r^2/2}.
\end{eqnarray}

Here, $l$ is an angular quantum number and $N_l$ is a normalization constant. Notice that the 
angular momentum operator commutes with  ${H_ \bot }$ for the SO(2) potential. Therefore, we  first get to $O(\epsilon^0)$: $\left\langle\{ \partial_s - i\tau L\}^2\right\rangle_0=\{\partial_s - i\tau l\}^2$, with $\left\langle \cdots \right\rangle_0$=$\left\langle \Phi_{l,0}(\theta)|\cdots |\Phi_{l,0}(\theta)\right\rangle$. 

All the other terms appearing in Eq. (\ref{teps}) yield averages of the form $\left\langle q^2_{1,2}\right\rangle_0 , \left\langle q_{1}q_2\partial_{1,2}\right\rangle_0, \left\langle q^2_{1(2)}\partial_{2(1)}\right\rangle_0$ (up to order ${\epsilon^{1/2}}$). When calculating these matrix elements using $|\Phi_l(\theta)>$, one obtains contributions where the angular momentum $l$ is changing by $\pm 1, \pm 2$ (related to, e.g., integrals of the form $\int_{0}^{2\pi} d\theta e^{i(l-m\pm 1)\theta}=2\pi \delta_{l-m\pm 1} $). Since there are no operators leading to transitions between transverse states with different angular momenta, all such contributions can be neglected, and we are left with: 

\begin{eqnarray}\label{effH1d}
{\hat{H}_{1D}}{\rm{ = (}}{\partial _s}{\rm{ - i}}\tau l{{\rm{)}}^2} + \frac{{{\rho ^2}}}{4},
\end{eqnarray}

\noindent which is, after the transformation $s=\phi/L$, the Hamiltonian of Eq. (1) in the main text. 

We stress that, as a result of the previous discussion and for this specific choice of SO(2) potential, the terms involving $\epsilon$ do not contribute even for a finite $\epsilon$. A similar analysis has been carried out in Refs. [3], and with a more rigorous mathematical presentation in Refs. [4,5]. We refer the interested reader to those articles for additional details as well as to the older study by Maraner [5].

Since the term $-i\tau l$ in Eq.~(\ref{effH1d}) appears now as a gauge field, it can be removed (for an infinite helix) by a unitary transformation. Alternatively, we could have started the analysis of Eq. (\ref{eq1}) by using a local orthogonal reference frame, where the metric tensor is diagonal. This can be obtained from the original frame by going to a locally rotating frame, where the normal and binormal vectors are transformed according to: 

\begin{equation}
\left( {\begin{array}{*{20}{c}}
\textbf{n}\\
\textbf{b}
\end{array}} \right) = \left( {\begin{array}{*{20}{c}}
{\cos \vartheta (s)}&{\sin \vartheta (s)}\\
{ - \sin \vartheta (s)}&{\cos \vartheta (s)}
\end{array}} \right)\left( {\begin{array}{*{20}{c}}
\textbf{N}\\
\textbf{B}
\end{array}} \right)
\end{equation}

\noindent The angle $\vartheta (s)= \int {ds'}\tau ({s'}) = \tau s$ for the case of a helical pathway with constant torsion and curvature. Using this orthogonal basis, which corresponds to a ``parallel transport'' frame, the term involving the angular momentum operator $\hat L$ does not appear, indicating that it can be removed by an appropriate change of frame. For this reason, it has not been considered in the main text, which only addresses an infinite helical system. 

As mentioned above, there are in principle many possible choices of the transverse confinement with increasing complexity, e.g., anisotropic potentials, anharmonic contributions, spin-dependent potentials (in the case in which an electromagnetic field is present), etc. These different {\textit {Ans\"atze}} deserve separate studies, which are beyond the scope of our present work. 

As a second example, we  address  another harmonic potential, which however breaks the SO(2) symmetry, and corresponds to a square well harmonic confinement as is typical of approaches to study quantum wells. In this case, the potential energy  is  ${V_\lambda }({q_1},{q_2}) = ({\lambda ^2_1}q_1^2 +{\lambda ^2_2} q_2^2)/2$, and the transversal ground state wave functions are given by: 
 \begin{equation}
     |\Phi_0(q_1,q_2)> = |\phi(q_1)>|\phi(q_2)>=M e^{-\lambda^2_1q^2_1/2} e^{-\lambda^2_1q^2_2/2}, 
 \end{equation}
where $M$ is a global normalization constant. Performing a similar analysis as before, we get in this case (for simplicity, we take $\lambda_1 = \lambda_2$): 

\begin{eqnarray}
  \hat H_{1D} =   \partial^2_s + \frac{\rho^2}{4} + 4\tau^2 g_1 + \epsilon g_2(\frac{\rho^2}{2})^2 (1 +6 \frac{\tau^2}{\rho^2}) +O(\epsilon^{3/2}),
  \label{eq15}
\end{eqnarray}

with $g_1= (\left\langle q_{1}\partial_{1}\right\rangle_0)^2\neq 0$ and $g_2=\left\langle q^2_{1}\right\rangle_0=\left\langle q^2_{2}\right\rangle_0\neq 0$. Notice that the term $O(\epsilon^1)$ is proportional to the mean square fluctuations along the transverse coordinates $q_1,q_2$. If only contributions $\sim \epsilon^0$ are kept, then there is, besides the curvature-dependent geometric potential, a torsion-dependent geometric potential: 

\begin{eqnarray}
  \hat H_{1D} =   \partial^2_s + \frac{\rho^2}{4} + 4\tau^2 g_1 + O(\epsilon),  
\end{eqnarray}

\noindent but the angular momentum term is absent, since the SO(2) symmetry is broken. The term proportional to $\tau^2$ arises from a contribution of the form $\left\langle q_1\partial_1 + q_2\partial_2 + 2 q_1q_2 \partial_1 \partial_2 \right\rangle_0$, which was originally part of $\hat L ^2$. 

As a last example, we consider a {square well with infinite walls}, i.e., 
$V(q_1,q_2)=0, \, \textrm{if} \; |q_{1,2}| \le \epsilon/2$ and $V(q_1,q_2)=\infty, \, \textrm{if} \; |q_{1,2}| > \epsilon$. The corresponding eigenfunctions are given by (assuming the same size of the confinement along the $q_1$ and $q_2$ directions): 

 \begin{equation}
     |\Phi_{,n}0(q_1,q_2)> = |\phi_n(q_1)>|\phi_n(q_2)>=\frac{2}{\epsilon} \cos^2{\frac{\pi}{2\epsilon}(2n+1)}, 
 \end{equation}

with the ground state corresponding to $ |\Phi_{0,n=0}(q_1,q_2)>=\frac{2}{\epsilon} \cos^2{\frac{\pi}{2\epsilon}}$. Along the same lines as in the previous two examples, we have to perform first the $\epsilon$-expansion and then carry out the projection on the ground state. The lowest order  terms which survive the  projection are $O(\epsilon^0)$, while the next higher order contributions are $O(\epsilon^{5/2})$ and will thus be left out:

\begin{eqnarray}
     \hat H_{1D} =   \partial^2_s + \frac{\rho^2}{4} + 2\tau^2\{  \left\langle q^2\partial^2\right\rangle_0 -   \left\langle q\partial\right\rangle_0 - (\left\langle q\partial\right\rangle_0)^2\} + O(\epsilon^{5/2}), 
\end{eqnarray}
 where we have already taken advantage of the fact that, e.g., $\left\langle q_1\partial_1\right\rangle_0=\left\langle q_2\partial_2\right\rangle_0$, etc. Using $\left\langle q^2\partial^2\right\rangle_0=1-\pi^2/6$ and  $\left\langle q\partial\right\rangle_0=-1$, we finally obtain : 

 \begin{eqnarray}
     \hat H_{1D} =   \partial^2_s + \frac{\rho^2}{4} + 2\tau^2 (1-\frac{\pi^2}{6}), 
\end{eqnarray}

which shows again a torsion-dependent contribution, which is in this case negative and thus reduces the influence of the curvature-dependent term.  

The different results when using the SO(2)  and the square-well potentials are not unphysical: they simply reflect the fact that different confinement potentials provide different symmetries of the boundaries in the neighborhood of the helical path and this manifests in different geometric potentials. Keeping the much smaller $O(\epsilon)$ contributions in the case of the square well with harmonic confinement or the $O(\epsilon^{5/2})$ terms in the square well with infinite walls will only add further smaller position-independent corrections to the quantum geometric potentials.

We stress that all the previously obtained corrections to the geometric potential are spin-independent and, for a given helical geometry (with a given pitch and radius), they will only provide a renormalization of  the energy origin in the electronic spectrum. This can be easely seen, e.g., by calculating the eigenvalues of the Hamiltonian in Eq. (5) in the main text, by including the additional corrections from Eq. (\ref{eq15}). The eigenvalues are given by:
\begin{eqnarray} 
\tilde{E}^{\kappa}_{\pm,l} = l^2  \pm \kappa l + \frac{1}{4}(1-\rho R) -
 4\tau^2 g_1 - \epsilon g_2(\frac{\rho^2}{2})^2 (1 +6 \frac{\tau^2}{\rho^2}), 
 \end{eqnarray}

\noindent which clearly shows that the spinor eigenvectors will be independent of $\epsilon$, and thus the obtained results on the charge and spin currents will not be affected by this energy corrections, since only the eigenvectors are used in their calculation. Moreover, since the geometric potentials commute with the spin rotation operator introduced in the main text, they do not play a fundamental role in the subsequent discussion that follows the introduction of the geometric spin-orbit coupling in the manuscript.

\ \\
\textbf{References}
\begin{enumerate}
\item Y.-L. Wang, M.-Y. Lai, F. Wang, H.-S. Zong, Y.-F. Chen, Phys, Rev. A \textbf{97}, 042108 (2018)
\item R. C. T. da Costa, Phys. Rev. A \textbf{23}, 1982 (1981).
\item K. Michaeli, R. Naaman, J. Phys. Chem C \textbf{123} 17043 (2019).
\item M. Geyer, R. Gutierrez, G. Cuniberti, J. Chem. Phys. \textbf{152}, 214105 (2020).
\item P Maraner, J. Phys. A: Math. Gen. 28, 2939 (1995). 

\end{enumerate}
}